\documentclass[manuscript, screen]{acmart}

\AtBeginDocument{%
  }

\acmJournal{TORS}   

\usepackage{booktabs}
\usepackage{hyperref}
 \usepackage{array} 
\usepackage{multirow}
\usepackage{colortbl} 
\usepackage{hhline} 
\usepackage{makecell} 
\usepackage{tabularx}
\usepackage{arydshln}
\usepackage{booktabs}
\usepackage{siunitx}
\usepackage{amsmath}
\usepackage{longtable} 
\usepackage{array} 
\usepackage{hyperref} 
\usepackage{tikz}
\usetikzlibrary{shapes, positioning, fit}
\usetikzlibrary{matrix, shapes, arrows.meta, positioning}
\usepackage{tikz}
\usetikzlibrary{positioning, shapes, arrows.meta, backgrounds}
\usepackage{multirow}
\usepackage{array}
\usepackage{xcolor}

\usepackage[most]{tcolorbox} 
\usepackage{enumitem}

\newcommand{\nm}[1]{\textcolor{red}{/* #1 (nm) */}}

\newcommand{\yashar}[1]{{\color{teal}[Yashar: \textbf{#1}]}}

\definecolor{myviolet}{rgb}{0.9,0.1,0.6}

\usepackage[most]{tcolorbox} 

\definecolor{bluegreen}{rgb}{0,0.5,0.5}
\newcommand{\updated}[1]{\textcolor{blue}{{\bf }{#1}{\bf}}}
\newcommand{\dquote}[1]{``#1''}

\usepackage{ifthen}

\begin{document}

\title{Agentic Personalized Fashion Recommendation in the Age of Generative AI: Challenges, Opportunities, and  Evaluation}

\author{Yashar Deldjoo}
\affiliation{%
  \institution{Polytechnic University of Bari}
  \country{Italy}}
\email{deldjooy@acm.org}
\orcid{1234-5678-9012}

\author{Nima Rafiee}
\affiliation{%
  \institution{Zalando}
  \city{Berlin}
  \country{Germany}}
\email{nima.rafiee@zalando.de}
\orcid{0000-0002-3193-9534}

\author{Mahdyar Ravanbakhsh}
\affiliation{%
  \institution{Zalando}
  \city{Berlin}
  \country{Germany}}
\email{mahdyar.ravanbakhsh@zalando.de}
\orcid{0000-0002-6456-867X}


\renewcommand{\shortauthors}{Deldjoo et al.}

\begin{abstract} 
Fashion recommender systems (FaRS) face distinct challenges due to rapid trend shifts, nuanced user preferences, intricate item-item compatibility, and the complex interplay among consumers, brands, and influencers. Traditional recommendation approaches, largely static and retrieval-focused, struggle to effectively capture these dynamic elements, leading to decreased user satisfaction and elevated return rates. This paper synthesizes both academic and industrial viewpoints to map the distinctive output space and stakeholder ecosystem of modern FaRS, identifying the complex interplay among users, brands, platforms, and influencers, and highlighting the unique data and modeling challenges that arise.

We outline a research agenda for industrial FaRS, centered on five representative scenarios spanning static queries, outfit composition, and multi-turn dialogue, and argue that mixed-modality refinement—the ability to combine image-based references (anchors) with nuanced textual constraints—is a particularly critical task for real-world deployment. To this end, we propose an Agentic Mixed-Modality Refinement (AMMR) pipeline, which fuses multimodal encoders with agentic LLM planners and dynamic retrieval, bridging the gap between expressive user intent and fast-changing fashion inventories. Our work shows that moving beyond static retrieval toward adaptive, generative, and stakeholder-aware systems is essential to satisfy the evolving expectations of fashion consumers and brands.
\end{abstract}




\maketitle

\section{Introduction}
The global fashion market—now surpassing \textbf{US \$2 trillion} in annual revenue—has become a crucible for the most demanding challenges in recommender system research~\cite{fortunebusinessinsights2023}. Fashion is a highly visual and emotionally driven domain, characterized by trends that can emerge and dissipate within days. Even a single misprediction may lead to costly reverse logistics, as dissatisfied customers frequently return ill-fitting garments. This volatility is compounded by a triadic ecosystem comprising consumers, brands, and e-commerce platforms, each with distinct and sometimes conflicting objectives~\cite{deldjoo2023review}. Additionally, influencers, sustainability advocates, and logistics partners further complicate the landscape. Recommendation pipelines designed for relatively stable domains—such as books, movies, or electronics—often prove inadequate when faced with the rapid trend fluctuations, subjective aesthetic preferences, and elevated return rates inherent to the fashion industry.

\begin{figure}[!t]
    \centering
    \includegraphics[width=0.60\linewidth]{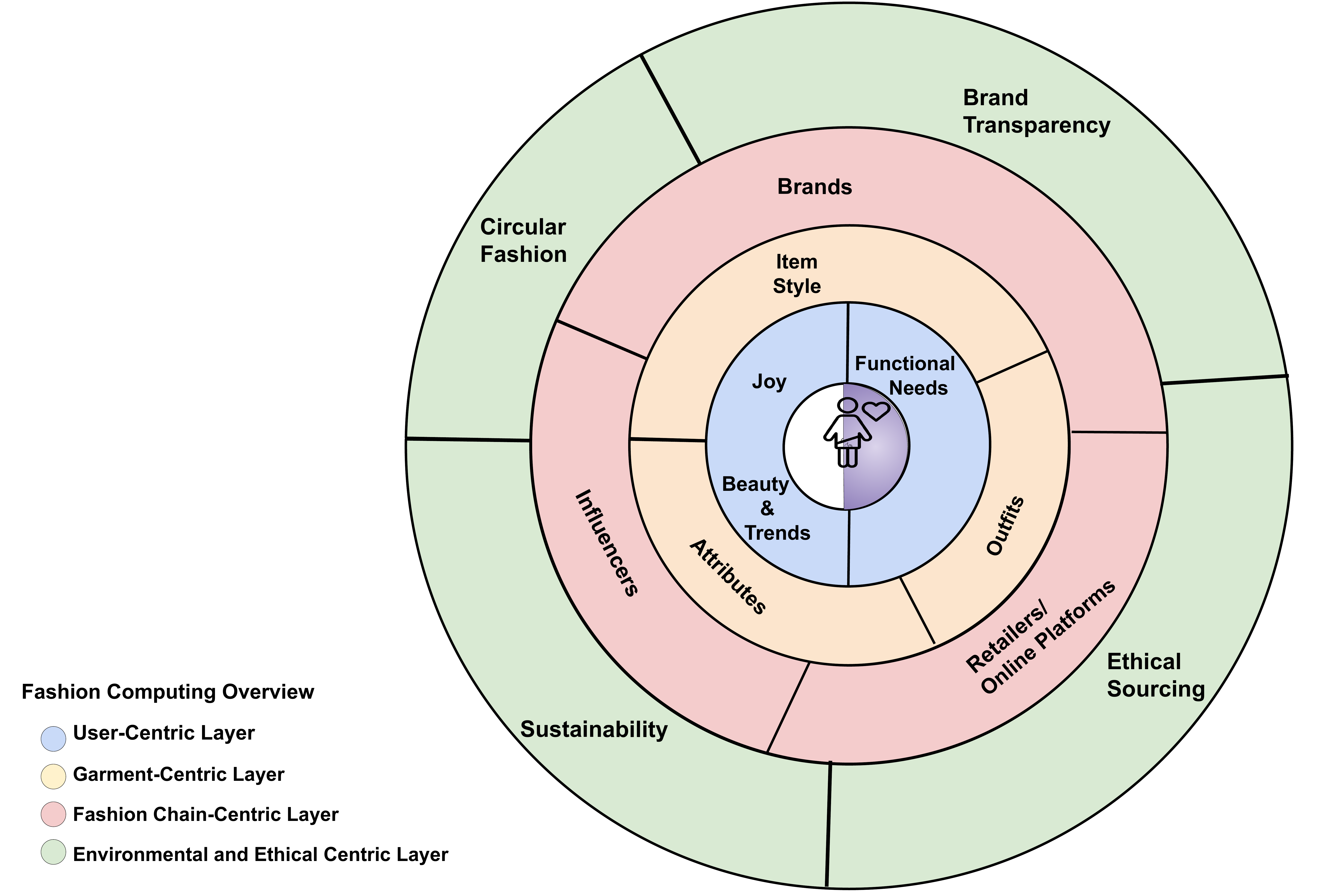}
    \caption{A layered conceptual model of fashion recommendation, with an inner focus on the user (joy, beauty and trends, functional needs), garment‐level attributes in the middle, chain‐level factors next, and outermost emphasis on sustainability and ethical considerations.}
    \label{fig:fashion_eco}
\end{figure}

Recent breakthroughs in large vision–language models (VLMs), diffusion-based image generators, and agentic large language models (LLMs) provide the technical foundation for a new generation of fashion recommenders that can \emph{perceive, reason, and act} in ways unattainable just a few years ago \cite{fashionclip,zalando2025fashionassistant,fashiongpt2025chen,10.1145/3626772.3657719}. Generative AI enables (i) rich multimodal grounding—for example, fusing an uploaded outfit photo with a textual request such as \dquote{formal-ish, K-pop inspired, under €200}; (ii) real-time data augmentation that alleviates cold-start pain by synthesizing embeddings for fresh inventory \cite{}; and (iii) dialog-based explanations that build user trust and reduce decision uncertainty, thereby lowering costly return rates and environmental impact.

This perspective paper has three main ambitions/contributions:

\begin{enumerate}[leftmargin=*]
\item \textbf{Raise collective awareness of the distinct attributes of fashion recommendation systems.}
\begin{itemize}
\item We compare and contrast fashion against music and general e-commerce along input–output granularity, trend velocity, item compatibility, stakeholder geometry, and return-cost externalities (see Table \ref{tab:comparison} in Sec. \ref{sec:diff}).
\item The latter holistic view can help to understand domain-specific difficulties—e.g., the lack of standard protocols for evaluating outfit-level compatibility or for capturing \dquote{style drift} over a season.
\end{itemize}

\item \textbf{Charting a research agenda around five promising fashion-RS tasks.} These include Static (Text), Static (Image), Mixed-Modality, Outfit Completion, and Multi-Turn Chat (see Table \ref{tab:scenario-summary}.) 

\item \textbf{Highlighting Mixed-Modality Refinement and proposing an agentic generative solution.} Specifically, we spotlight mixed-modality refinement as a particularly critical task, demonstrating why existing retrieval-only pipelines fail to address core challenges such as unseen attributes, rapid trend shifts, and compositional queries. We propose \textbf{AMMR (Agentic Mixed-Modality Refinement)}, a powerful generative pipeline leveraging multimodal encoders, dynamic query composition, and an LLM-based agentic planner to deliver fast, accurate, and constraint-aware recommendations (\S\ref{sec:ammr}), see as well~\citet{maragheh2025future} for a good frame of reference.


\end{enumerate}

Figure~\ref{fig:fashion_eco} summarizes the core elements of a Fashion Recommender System (FaRS), structured around four interconnected macro levels. First, the \emph{user level} addresses diverse motivations such as aesthetic preferences, trend-following, or practical needs like comfort and functionality. Second, the \emph{garment level} covers specific item attributes (color, design, fabric) and outfit compatibility. Third, the \emph{fashion chain level} encompasses key stakeholders including brands, retailers, platforms, and influential actors who collectively shape consumer trends through events and social media. Finally, the \emph{supply-chain and ethical level} emphasizes increasingly crucial concerns such as sustainability, circular fashion practices, and responsible sourcing, reflecting broader societal values~\cite{Eldemerdash2023FashionRS,10.1145/3523227.3547377}.

\section{The Output Space of FaRS}
\label{sec:output-space}

FaRS produce a range of outputs, some 
directly visible to end users and others primarily used internally by the 
system. In this section, we discuss these outputs and the diverse array 
of actual fashion products FaRS handle.

\subsection{User-facing vs.\ Hidden Outputs}
FaRS mainly matches users with \emph{fashion products}, but can also produce a variety of additional outputs. We can broadly categorize these as \emph{user-facing} (visible to end users) versus \emph{non-user-facing} (mostly internal to the system). Recognizing both categories clarifies \emph{what} FaRS delivers (items, explanations, etc.) and \emph{how} the underlying mechanisms operate (embeddings, data augmentations, etc.). Below, we provide further details on these categories.
\begin{itemize}
    \item \textbf{Recommendations (Single Items or Outfits).}  In the fashion domain, recommendations span a wide range of products that differ by \textit{type} (e.g., garments, accessories) and \textit{grouping} (e.g., individual items or complete outfits). To maintain coherence, FaRS must account for multiple relationships—such as \emph{item--item} and \emph{user--item} interactions—while also considering user profiles, contextual factors and different stakeholder objectives.
    \item \textbf{Styling Tips and Explanations.} In the fashion domain, users—especially women, who typically face a broader range of product choices—often experience significant uncertainty about which recommended items will suit them best \cite{Shirkhani2023StudyOA}. For example, many women often seek the opinion of a friend before committing to a new look. This uncertainty is due to many factors at play -- such as whether the items match their personal style, harmonize well to complete a wardrobe, or are truly on-trend. Providing clear, \textbf{personalized} and \textbf{expert-driven} explanations can help alleviate these doubts, build trust, and encourage users to experiment confidently with new looks.

    \item \textbf{Conversational or Generative Content.} Large Language Models 
    (LLMs) can generate multi-turn dialogues or textual narratives about 
    the style of a target user. They may also create dynamic descriptions or AR-based 
    outfit previews that go beyond static product feeds.
\end{itemize}

Internally, FaRS rely on \emph{hidden outputs}, not typically shown to users. These are \textbf{intermediate artifacts} that are central to how modern FaRS process data and rank results. These outputs typically remain hidden from the user, although they can sometimes be optionally displayed—such as revealing the reasoning process opted by the model, such as:
\begin{itemize}
    \item \textbf{Latent Embeddings and Feature Maps.} Neural encoders (e.g., 
    CLIP-like models) transform item images and textual attributes into 
    compact representations that capture similarity and style compatibility \cite{fashionclip, chia2022contrastive}.
    \item \textbf{Intermediate Candidate Sets.}A subset of items retrieved at an earlier stage (e.g., from a vector database or knowledge base) in systems such as retrieval augmented generation (RAG).
    
\end{itemize}

\subsection{Product Diversity in FaRS}
While the previous sections describe how the FaRS structure \emph{outputs} in general, actual recommendations in the fashion domain extend beyond garments (clothing items) and also include other types of products such as:
\begin{itemize}
    \item \textbf{Apparel and Accessories:} Shirts, dresses, pants, outerwear, 
    handbags, jewelry, and so on.
    \item \textbf{Footwear:} Shoes, boots, and sneakers curated for style or 
    comfort.
    \item \textbf{Beauty and Home Decor:} Cosmetics or brand-aligned decor 
    (e.g., IKEA collaborations) relevant to personal style.
    \item \textbf{Complete Outfits or Capsule Wardrobes:} A multi-item set 
    matching a user’s silhouette, color palette, or intended occasion.
\end{itemize}
one main challenge in the fashion domain is controlling different types of relationships beyond the user-item interaction (as seen in classical RS). See the next section for more details.

\begin{table*}[!t]
\centering
\caption{Comparison Across Fashion, Music, and General E-Commerce}
\label{tab:comparison}
\setlength{\tabcolsep}{3pt} 
\renewcommand{\arraystretch}{1.5} 
\small 
\begin{tabular}{@{}p{3.1cm}p{3.5cm}p{4.0cm}p{5.0cm}@{}}
\toprule
\textbf{Dimension} & \textbf{Fashion} & \textbf{Music} & \textbf{General E-Commerce}\\ 
\midrule
\rowcolor[HTML]{FFFFFF} 
\textbf{User Goals} & 
Express style; fit \& function & 
Discover new songs/artists &
Find desired items (price, convenience)\\
\rowcolor[HTML]{F9F9F9} 
\textbf{Visual Relevance} & 
Extremely high (color, texture, shape) &
Low (audio-based domain) &
Mostly textual/feature-based (reviews, specs)\\
\rowcolor[HTML]{FFFFFF} 
\textbf{Trend Sensitivity} &
Very high; seasonal cycles &
Moderate; cultural/music trends &
Moderate; trending products or seasonal deals\\
\rowcolor[HTML]{F9F9F9} 
\textbf{Multi-Stakeholder} &
Brands, influencers, stylists &
Labels, artists, streaming platforms &
Retailers, manufacturers, affiliates\\
\rowcolor[HTML]{FFFFFF} 
\textbf{Item Compatibility} &
Coherent outfits &
Playlist vibe/genre coherence &
Lower synergy demands\\
\rowcolor[HTML]{F9F9F9} 
\textbf{Returns \& Fit} &
High (size/fit issues) &
N/A (digital goods) &
Moderate; item dissatisfaction\\
\rowcolor[HTML]{FFFFFF} 
\textbf{Brand Loyalty \& Identity} &
Strong brand resonance &
Some label/platform loyalty &
Varies; cost-driven or brand-based\\
\rowcolor[HTML]{F9F9F9} 
\textbf{Subjectivity} &
Highly personal, aesthetic &
Personal taste + mainstream popularity &
More functional or specs-driven\\
\bottomrule
\end{tabular}
\vspace{1em}
\end{table*}

\section{What Makes Fashion Different?}
\label{sec:diff}
Fashion recommender systems differ substantially from other general e‐commerce applications. Table~\ref{tab:comparison} offers a concise overview of the contrasts between fashion, music domain, and generic e-commerce, which we expand on below.

\subsection*{Multifaceted Intertwined Relationships}
Unlike many other verticals (e.g., music, books, electronics), fashion recommendation systems must account for multiple, intertwined relationships:

\begin{itemize}[leftmargin=1.3em]
    \item \textit{Item--User relationships} (e.g., aligning with a user’s style preferences for casual or formal wear).
    \item \textit{Item--Item relationships} (e.g., color or fabric compatibility, as well as functional considerations like pairing a warm sweater with waterproof boots).
    \item \textit{Body or Face--Item relationships} (e.g., clothing or makeup that complements a user’s skin tone, body shape, or facial features).
\end{itemize}
This broad \dquote{anchor} makes user modeling both diverse and challenging \cite{Chakraborty2021FashionRS, li2020fashion}. Subjective style and cultural backgrounds heavily influence acceptance of recommended products, and balancing these multifaceted factors remains a challenge for FaRS \cite{matzen2017streetstyle}.

\subsection*{Multi-Stakeholder Complexity}
In the fashion domain, the business ecosystem is commonly described as a \textbf{three-sided market} involving:

\begin{enumerate}
    \item \textbf{Consumers} (e.g., Alex) who want personalized, on-trend, and wallet-friendly products that fit their size and style;
    \item \textbf{Brands} (e.g., Nike) that supply the inventory and strive for higher visibility, sales, and a positive return on investment (ROI) \cite{friedlob1996understanding, Mersni2024WhenGM};
    \item \textbf{Platforms} (e.g., Zalando) that connect brands and consumers, aiming to keep users engaged, protect brand partnerships, and meet revenue objectives.
\end{enumerate}

Although other e-commerce sectors also feature multiple stakeholders, \textbf{brand identity} and \textbf{brand family structures} (such as Nike Sport vs. Nike Essentials) are especially pivotal in fashion. Brands often require datasets with explicit brand labeling---both to \textbf{enforce producer fairness} (i.e., ensuring equitable visibility among different brands) and to maintain consistent brand images. Because of this, the availability and quality of brand-labeled data become crucial when designing fair and effective fashion recommendation pipelines \cite{Chakraborty2021FashionRS}.

Beyond these three primary stakeholders, \textbf{influencers} act as powerful catalysts in the fashion space \cite{nurfadila2020impact}. In some contexts, influencers might be considered a full-fledged 4\textsuperscript{th} stakeholder, but more often they amplify or accelerate consumer awareness, shaping brand preferences and intensifying marketplace dynamics. Platforms and brands thus devote significant resources to coordinating with these trendsetters, so that product assortments and marketing messages stay timely and culturally relevant. Additionally, sustainability, pollution control, and social responsibility are other crucial factors and stakeholders within the fashion ecosystem. 

\subsection*{High Visual and Aesthetic Demands}

\textbf{Fashion} is intensely visual, with color palettes, textures, and designs needing careful coordination across \textbf{body} (e.g., tops, pants, jackets, bags). This leads to sophisticated item compatibility requirements: a sleek blazer might clash with neon pants, even if both are individually popular. By contrast, \textbf{music} involves audio similarity and playlist coherence, but combining two slightly mismatched songs may still yield an acceptable listening experience. In \textbf{general e-commerce} (e.g., electronics), product synergy often matters less—consumers typically buy items in isolation.

\paragraph{Example:} 
A user searching for a "bomber outfit" might require a jacket, top, jeans, and shoes that align in texture and color. To meet this need, FaRS must recognize that "bomber" refers to a specific jacket style, typically waist-length with a fitted waistband and cuffs, and suggest tops and jeans that match the looser, retro-inspired design.

\subsection*{Intense Trend Sensitivity and Influencer Impact}
Fashion experiences rapid, often short-lived (seasonal and cultural) swings. A jacket popular this winter may be outdated next year, and social media influencers or celebrities can spark micro-trends on TikTok or Instagram overnight. Beyond influencers, \textbf{major events} such as a blockbuster movie or a TV series release can instantly elevate certain styles—think of a show featuring Victorian costumes, causing lace-up boots and puffed sleeves to trend. To remain effective, FaRS must adapt to these spikes and retire stale trends just as quickly, blending real-time social signals with historical data \cite{chakraborty2020predicting}. \textbf{Music} trends also evolve, but they typically cycle at a slower pace (e.g., certain genres gain or lose popularity over the years). In \textbf{general e-commerce}, while some products have seasonal peaks (e.g., holiday gifts), the domain overall sees fewer drastic style changes.


\paragraph{Example:} 
The video of a \dquote{capsule wardrobe} of an influence might suddenly boost sales for beige trench coats, forcing the platform to re-prioritize recommendations. This degree of immediate, visual trend disruption is rarer in music or mundane product categories such as office supplies. 


\subsection*{From Mood to Long-Term Investment}

Unlike music, where people easily switch playlists based on momentary emotion, \textbf{fashion purchases} tend to be costlier, involve physical items, and remain part of a wardrobe for months or years. \textbf{Short-term mood} (e.g., wanting something playful for a party) merges with \textbf{long-term expectations} (e.g., does it match other items in the closet? Will it still be wearable next season?). This makes fashion decisions more deliberative.

\begin{itemize}
    \item \textbf{Music}: Low barrier to change—no physical ownership, minimal risk.
    \item \textbf{Fashion}: Higher price point, physical space constraints, and potential for buyer’s remorse if the style or size turns out unsuitable.
    \item \textbf{General E-commerce}: Often driven by immediate need (e.g., a blender) or cost/function trade-offs, with less emphasis on stylistic longevity.
\end{itemize}


\subsection*{Style, Size, and Fit}

In fashion, three important item-level properties are style, size, and fit. Although users have preferences for each of these properties, it is the items themselves (or the outfits they form) that come with specific style, size, and fit characteristics.

\textbf{Style (item or  item-item level)} is an \emph{aesthetic property} that arises in both user--item and item--item relationships. On the user--item side, it reflects an individual’s aesthetic preferences, such as favoring minimalist or avant-garde designs. On the item--item side, it captures the coherence of multiple pieces when combined into an outfit. A single garment might be perfectly aligned with a user’s taste, but the overall look depends on how its style interacts with other items in a wardrobe or ensemble.

\textbf{Size (item-level )} is a \emph{geometrical property} at the item level, typically denoted by labels such as “Small,” “Medium,” or “Large.” It does not directly depend on who wears the garment, though discrepancies between a user’s assumed size and the actual cut of an item often lead to returns.

\textbf{Fit (body-item, or user-item)} is also a \emph{geometrical property}, yet it connects more closely to the user--item relationship (and, by extension, the user’s body). Common descriptors like “slim,” “regular,” or “loose” indicate how a garment is intended to drape on a wearer’s figure. Even if a user’s preferred style and size match an item, it may be returned if the fit does not align with the user’s comfort or body shape. ``Size'' and ``fit'' together are important for reducing return rates because ensuring that a garment not only matches a user's aesthetic but also fits their size preferences and body shape well minimizes misfit purchases \cite{roland_size_fit, amazon2024ai}.










\subsection*{Brand Loyalty, and ROI}
 Brand identity holds special importance in fashion, often outweighing cost and convenience \cite{Mersni2024WhenGM}. Some users buy exclusively from eco-friendly or luxury labels, reflecting personal values or status. From the perspective of brands, (\textbf{ROI}) typically refers to the \textit{ratio between net profit and the cost of marketing or inventory investments} over a given timeframe. For instance, if a brand spends \$10,000 on an influencer-driven campaign and sees a net profit increase of \$30,000 from resulting sales, the ROI is 3:1 \cite{friedlob1996understanding}. FaRS may prioritize brand-related placements to improve such ROI metrics, which must be balanced against user-centric relevance \cite{Jannach2023ASO}.

\textbf{Music} also sees label loyalty (e.g., fans of specific record companies) but far less intensely than fashion’s brand-driven culture. \textbf{General e-commerce} might feature brand-oriented shoppers—yet typically, functional specs or price remain paramount for big-ticket items like TVs or washing machines. In fashion, brand narrative and aesthetics often supersede practical criteria.

\paragraph{Example:} 
A shopper fixated on sustainable fashion might pay a premium for a brand’s limited-edition organic cotton line, ignoring cheaper alternatives. In general e-commerce, cost comparisons and feature lists usually dominate, overshadowing brand narrative.

\begin{figure}[!t]
  \centering
  \includegraphics[width=0.56\columnwidth]{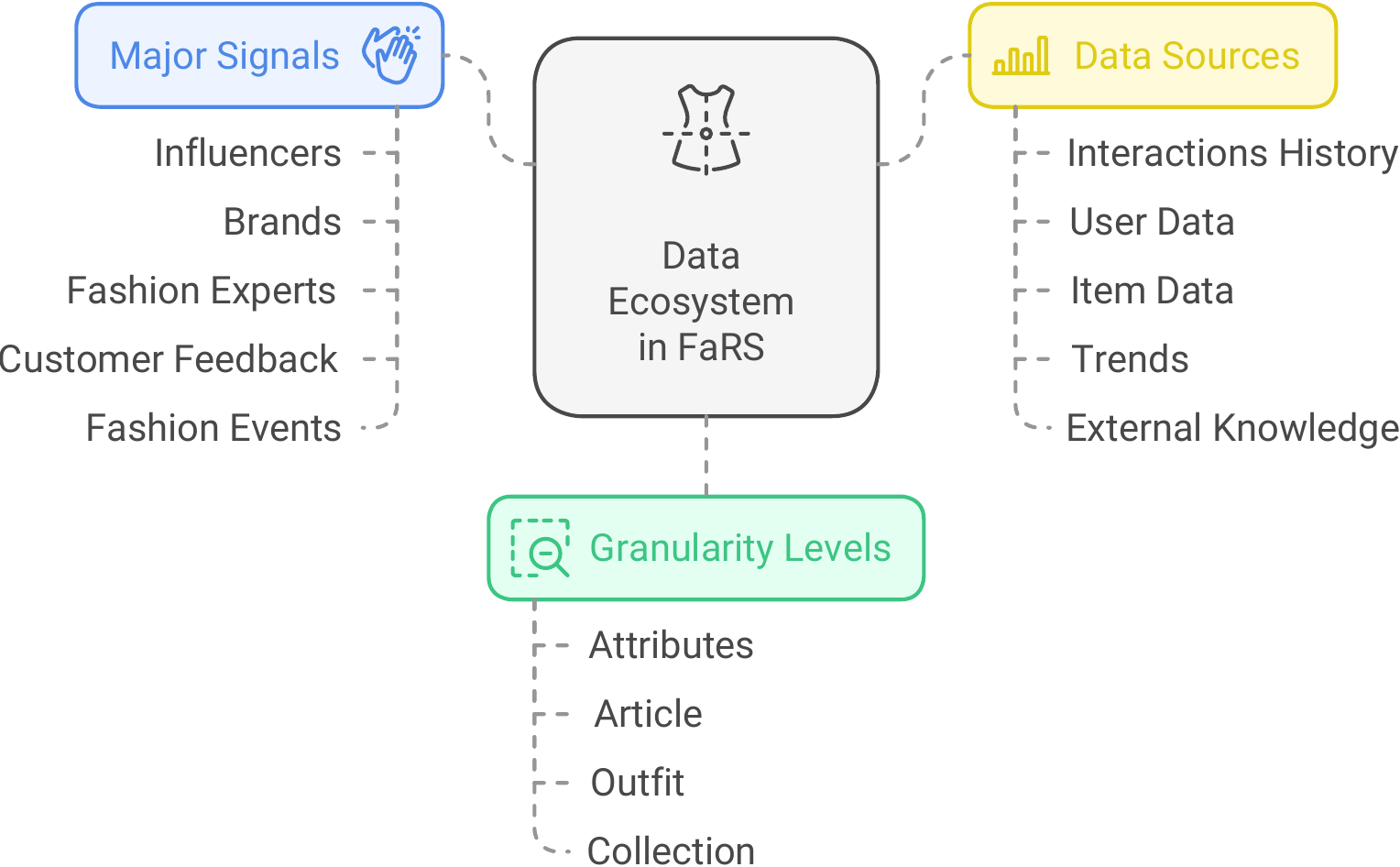}
\caption{FaRS data ecosystem.}
\label{fig:data_ecosystem_1}
\end{figure}

\medskip
\noindent
While conversational tool is useful for iterative suggestions, but feedback can also be gathered through visual interactive/gamified tools. This allows users to adjust preferences such as material, color, or brand, and see the recommendations update instantly. Overall, as outlined in Table \ref{tab:scenario-summary}, these scenarios illustrate how each variation in input, composition level, and interaction style brings distinct modeling challenges. The next sections will discuss how FaRS can leverage diverse data signals (cf Sec.~\ref{sec:data-ecosystem}), handle outfit-level composition, and integrate modern generative AI techniques to meet the evolving demands of real-world fashion platforms.

\section{Data Ecosystem and Fashion-Embedding Models}
\label{sec:data-ecosystem}
\subsection{The Data, input and Output}

Figure~\ref{fig:data_ecosystem_1} illustrates a conceptual overview of the \emph{data ecosystem} in Fashion Recommender Systems (FaRS). Central to this ecosystem is the integration of various data dimensions, each crucial to providing effective and responsive fashion recommendations. Specifically, the data ecosystem comprises three interconnected pillars: Major Signals, Data Sources, and Granularity Levels, which collectively shape recommendation outcomes.

\textbf{Major Signals (blue node).} These signals represent dynamic factors frequently influencing trends, user preferences, and brand strategies. FaRS must continuously monitor and adapt to these signals to maintain recommendation relevance:
\begin{itemize}[leftmargin=1.5em]
    \item \emph{Influencers} who trigger rapid micro-trends via social media.
    \item \emph{Brands}, publishing seasonal lookbooks and curated collections, influencing consumer tastes and platform merchandising.
    \item \emph{Fashion experts} and editorial teams, shaping broader fashion narratives.
    \item Direct \emph{customer feedback}, refining recommendation accuracy through real user insights.
    \item Major \emph{fashion events} (runway shows, fashion weeks), setting style benchmarks and seasonal expectations.
\end{itemize}

\textbf{Data Sources (yellow node).} The effectiveness of FaRS significantly depends on high-quality, structured and unstructured data inputs, ensuring accurate, timely, and personalized recommendations:
\begin{itemize}[leftmargin=1.5em]
    \item \emph{Interaction History}, including user clicks, purchases, and returns, revealing immediate preferences.
    \item \emph{User Data}, such as demographic information, explicit style profiles, and detailed textual feedback (e.g., product reviews stating ``the sleeves are too short''), enhancing personalization precision.
    \item \emph{Item Data}, sourced from comprehensive product catalogs describing visual and textual item attributes (color, fabric, brand).
    \item \emph{Trend Data}, real-time or near-real-time signals capturing ephemeral shifts in fashion preferences.
    \item \emph{External Knowledge}, comprising domain ontologies for clothing types, brand-specific constraints, and third-party fashion API integrations.
\end{itemize}

\textbf{Granularity Levels (green node).} Fashion recommendations occur at multiple levels of detail, accommodating diverse user intents and scenarios. FaRS must effectively manage these granularity levels to fulfill varying demands:
\begin{itemize}[leftmargin=1.5em]
    \item \emph{Attributes}, enabling users to specify fine-grained details such as colors, patterns, or materials.
    \item Individual \emph{Articles}, addressing straightforward user queries or searches for specific fashion items (e.g., a certain T-shirt or shoes).
    \item \emph{Outfits}, combining multiple articles into coherent looks, satisfying compatibility, occasion suitability, and stylistic coherence.
    \item \emph{Collections}, including comprehensive seasonal lineups or curated capsule wardrobes tailored to broader lifestyle needs or fashion trends.
\end{itemize}

\begin{table*}[t]
\centering
\caption{Summary of Example FaRS Scenarios, Their Inputs/Outputs, and Key Challenges.}
\label{tab:scenario-summary}
\small
\begin{tabular}{l p{4.0cm} p{2.0cm} p{5cm}}
\toprule
\textbf{Scenario} & \textbf{Input--Output} & \textbf{Output space} & \textbf{Representative Challenges} \\
\midrule
\textbf{1 Static (Text)}    & Text query \newline (e.g., ``gothic T-shirt'') $\rightarrow$  item list & Single item & Bridging semantic gaps; reconciling user vs.\ brand goals; rapid trend shifts \cite{chia2022contrastive, Li2022MultiStakeholder, chakraborty2020predicting, Shin8966228}\\
\rowcolor[HTML]{F9F9F9} 
\rowcolor[HTML]{F9F9F9}
\textbf{2 Static (Image)} & User-uploaded image $\rightarrow$ item list & Single item & Defining similarity; subjective nuances; robust vision pipelines \cite{costomizedOutfit, varma2023villa}\\
\rowcolor{yellow!20}{\textbf{3 Mixed-Modality}} & Image + text note $\rightarrow$ item list & Single item & Multi-modal fusion; style generalization; personalization re-ranking \cite{mmfrec2023, chen2025multifactorial, wang2024mccp}\\
\rowcolor[HTML]{F9F9F9}

\textbf{4 Outfit Completion}& Photo of user's item $\rightarrow$ matching top/bottom & Outfit-level & Item--item compatibility; user-specific fit; limited multi-item logs \cite{vto_outfit_survey,fashion_AR,costomizedOutfit}\\
\textbf{5 Multi-Turn Chat}  & Text prompts across multiple turns $\rightarrow$ dynamic recs & Outfit- or single-item + textual explanation& Fashion-aware dialogue; iterative re-ranking; multi-agent integration \cite{kasra2025best_llm_ever, sayana2024beyond, fang2024multi}\\
\bottomrule
\end{tabular}
\end{table*}
By unifying these three interconnected pillars—Major Signals, Data Sources, and Granularity Levels—FaRS can adeptly respond to both general and highly specific fashion queries. This holistic approach naturally bridges into the next critical component: the \emph{outfit-embedding model}, detailed subsequently, which operationalizes this rich data ecosystem into actionable, coherent, and personalized fashion recommendations.

Modern fashion search increasingly demands the ability to \emph{refine} queries using both visual and textual input. Users may upload a photo to capture a desired silhouette or style, but then express additional preferences---such as changing the color, adding a belt, or specifying a particular feature like pockets---in natural language. As described in~\cite{9857242_baldrati,empirical_odyssey,Saito2023Pic2WordMP,9710082_liu,Radford2021LearningTV}, this \textbf{mixed-modality refinement} paradigm enables more expressive and natural interaction:

\section{Mixed-Modality Refinement in Fashion}

Table~\ref{tab:scenario-summary} summarizes five key recommendation tasks typically encountered on major fashion platforms such as Zalando. Due to its significant business value, enabling expressive user interactions and the substantial open research space, this work particularly focuses on \textbf{Mixed-Modality Refinement (MMR)}, leaving the remaining tasks as promising directions for future exploration.

\subsection{Motivation to MMR.} In modern fashion search, consumers increasingly want to combine \dquote{search by look} with \dquote{search by specification.} For instance, an uploaded photo might capture the overall silhouette or style a user loves, yet the user might still need to change or adjust specific details—color, length, or presence of certain features (e.g., pockets, collars, etc.). In classical \textbf{single-modal} search, these nuanced constraints are easy to miss. A purely text-based approach struggles to convey the exact visual style the user wants to keep. Conversely, a purely image-based approach may not capture the user’s new demands (\textit{\dquote{add a pocket,}} \textit{\dquote{shorten the hem,}} \textit{\dquote{switch to black suede}}). The \emph{mixed-modality refinement} enables a highly expressive and interactive search experience, allowing users to articulate requests such as 

\begin{quote}
    ``I love everything about this \includegraphics[height=1.5em]{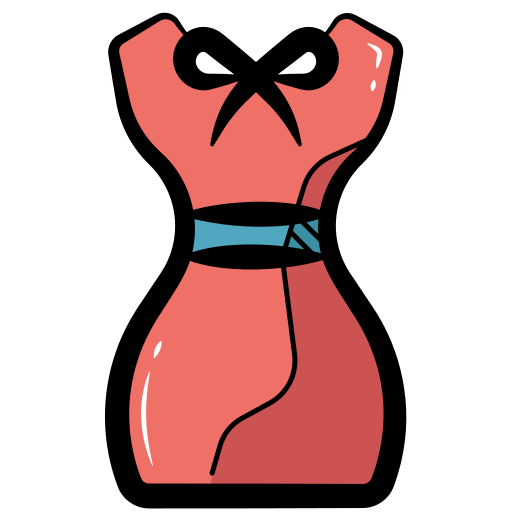}---except I’d like it in a darker color and with a belt.''
\end{quote}

or, more generally,
\begin{quote}
    ``More like this \includegraphics[height=1.5em]{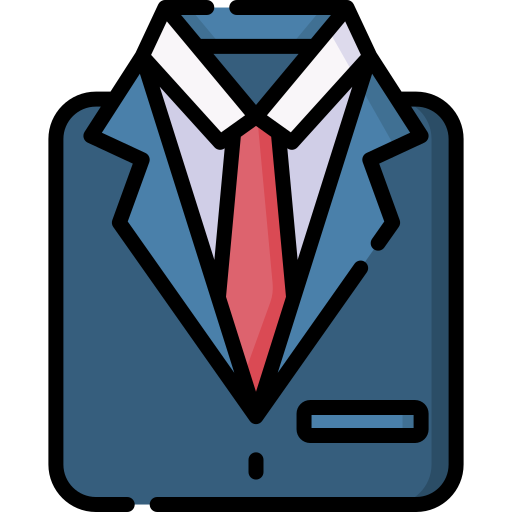}, but change X''
\end{quote}

\noindent Overall, it is fair to state that in the \textbf{fashion} domain, visual and textual attributes interact in subtle ways. Purely image-based search ignores specific requests, while text-based search struggles to convey nuanced visual style. Mixed-modality refinement thus aims to retrieve items that \emph{preserve} the user's reference style while satisfying explicit textual modifications. Overall, the objective in \dquote{Mixed-Modality Refinement} is to retrieve catalog items that not only resemble the reference image but also satisfy the specific modification described by the user. Unlike traditional keyword- or tag-based search, this paradigm supports iterative, compositional queries that more naturally capture the user’s intent and foster creative exploration \cite{9857242_baldrati,empirical_odyssey,Saito2023Pic2WordMP,9710082_liu,Radford2021LearningTV}.\\

\noindent \noindent \textbf{Production Requirements.}
An industrial-grade mixed-modality engine must

\begin{itemize}[leftmargin=1.2em]
  \item \textbf{Respond in real time} (low-latency high-throughput).\vspace{2pt}
  \item  \textbf{Scale} to tens of millions of items and daily queries.\vspace{2pt}
  \item \textbf{Combine accuracy and control}: retrieved results must satisfy both the visual anchor and \emph{all} textual constraints.\vspace{2pt}
  \item \textbf{Generalise to newly emerged styles and attributes} that never appeared during initial training.\vspace{2pt} 
  \item \textbf{Remain sensitive to fine-grained user intent} so that even subtle edits (e.g., sleeve length, pocket shape) lead to perceptibly better recommendations and a positive customer experience. 
\end{itemize}


\subsection{Retrieval-Only Mixed-Modality Search}
\label{sec:retrieval_only}
Industrial fashion platforms still favour \emph{retrieval-only}
pipelines because they plug directly into high-throughput ANN indices
and keep median latency below 200 ms.  Broadly, these pipelines fall
into two stages: a \emph{universal embedding} that is pre-computed
offline, and an \emph{online composition step} that injects the user’s
mixed-modal query.  We review both in turn.

\noindent\textbf{Universal Embedding Baseline.}
Let a vision–language backbone (e.g.\ CLIP) embed every catalogue
item $x$ once,
$\phi(x)\!\in\!\mathbb{R}^d$, and let it embed at query time

\begin{itemize}[leftmargin=1.5em, nosep]
  \item the user’s reference images $\mathcal I$,
  \item the textual modification \textit{text},
\end{itemize}

Nearest-neighbour search on cosine similarity
retrieves candidates in the \emph{same} space.

\paragraph{Advantages.}
\begin{itemize}[leftmargin=1.8em, nosep]
  \item \textbf{Simplicity.} One embedding pipeline for all categories
        and all query types.
  \item \textbf{Scalability.} Once $\phi(\cdot)$ is indexed, billions
        of similarity probes are handled by off-the-shelf ANN tools
        such as Faiss, ScaNN, or Annoy.
  \item \textbf{Zero/Few-Shot Robustness.} Large foundations often
        cope with unseen classes “for free.”
\end{itemize}

\begin{figure}[!h]
    \centering
    \includegraphics[width=0.93\linewidth]{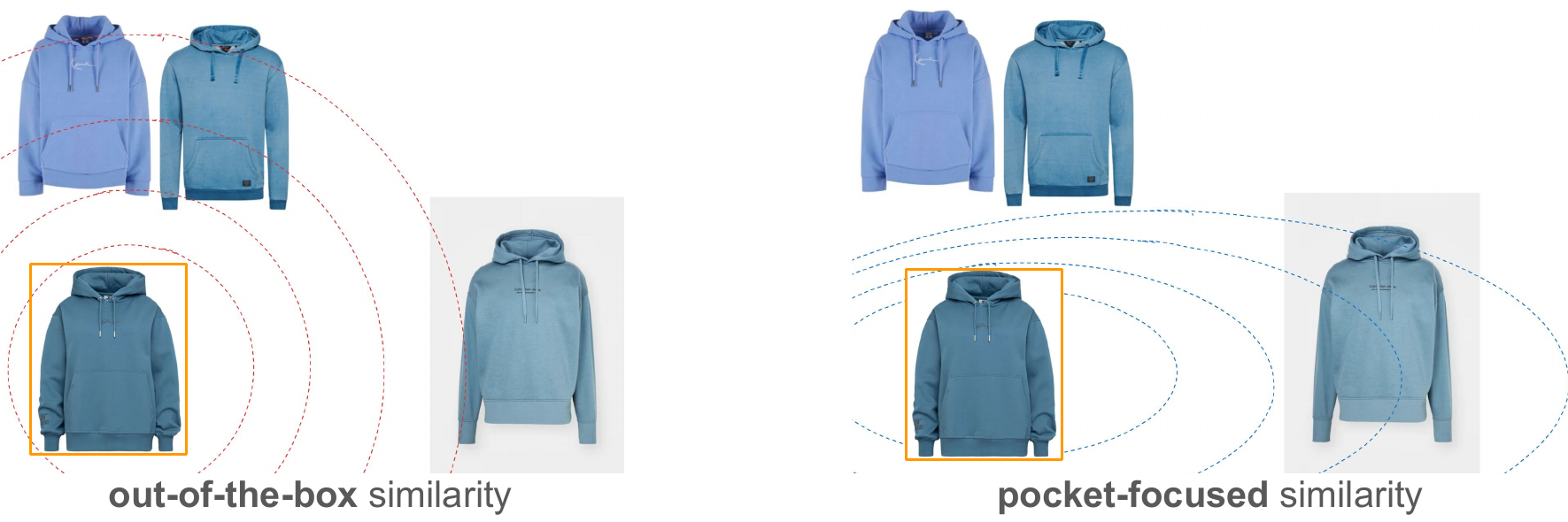}
 \caption{Illustration of dynamically reshaping similarity. Given a visual cue (orange box, `blue hoodie with a pocket'') and a refinement request (`without a pocket''), traditional methods struggle. By emphasizing the pocket attribute, the customized neighborhood better captures user intent.}
    \label{fig:costum_sim}
\end{figure}
Current universal embeddings struggle with fine-grained fashion attributes, imbalanced category distributions, and rapidly evolving trends that demand frequent retraining. Rare details and new styles are often overlooked, while static embeddings fail to support compositional queries like negation or nuanced comparisons. These limitations highlight the need for more adaptive and semantically expressive representations.


Figure~\ref{fig:costum_sim} makes the problem with fine-grained fashion attributes tangible with a hoodie
query. The orange-framed anchor garment has a kangaroo pocket. On the left, in the out-of-the-box space (red contours), color and coarse category dominate, causing pocketed hoodies to cluster near the anchor, while a pocket-free candidate is pushed to the periphery
\emph{Left:} in the \textit{out-of-the-box} space (red contours) colour
and coarse category dominate, so pocketed hoodies gravitate towards
the anchor while an actually pocket-free candidate is pushed to the
periphery.
\emph{Right:} when we “tilt’’ the metric space to emphasise the binary
\textsc{has-pocket} attribute (blue contours) the situation inverts:
items that share the \emph{absence} of a pocket move inward, and the
previously top-ranked—but pocketed—hoodies slide outward.
Only this re-weighted view makes a refinement such as
“same hoodie, but without a pocket’’ (or conversely “remove the pocket’’)
meaningful to the retrieval engine.

\noindent\textbf{Query-Time Composition Operators.}
To overcome the rigidity of a universal metric, retrieval-only systems
learn a small \emph{composition operator} $g_\theta$ that fuses the
image embedding $v=f_v(\mathcal I)$ and the text embedding
$t=f_t(\text{text}) \in\mathbb{R}^d$ into a query vector
\[
  q \;=\; g_\theta(v,t),
  \qquad
  \text{rank}(x) \;=\; \operatorname{sim}\bigl(q,\phi(x)\bigr).
\]
Two influential operator families are outlined below.

\subsubsection*{$\Delta$-Shift on Disentangled Sub-Spaces}

Make the vision encoder partition its latent vector into orthogonal
slices—colour ($c$), material ($m$), silhouette ($s$), …—
$
  v=[\,c\!\parallel m\!\parallel s\!\parallel\!\cdots\,]
$
\cite{ma2019learning,wan2023dual}.  
A text-conditioned delta
$\Delta = tW$
is \emph{added only to the relevant slice(s)}:
\[
  q_{\Delta}= [\,c+\Delta_c \parallel m+\Delta_m \parallel s+\Delta_s
               \parallel\!\cdots\,].
\]
Triplet-ranking loss plus slice-orthogonality
enables attribute-specific control while keeping ANN latency intact.

\subsubsection*{Residual / Gating Masks (TIRG-Style)}

Keep $v$ monolithic and let the text decide what to modify
\cite{vo2019composing}:
\[
  q_{\textsc{TIRG}}
  = W_0\,v \;+\; \bigl(M(t)\odot v\bigr).
\]
Dense masks alter the whole vector; sparse masks approximate the
$\Delta$-shift effect \emph{without} explicit disentanglement.

\paragraph{After-market variants.}
Attention fusion (MAAF, AACL)\cite{dodds2020modality,tian2023fashion}, graph-smoothed $\Delta$ vectors (RTIC) \cite{shin2021rtic}, or

auto-encoding manifolds (ComposeAE) \cite{anwaar2021compositional} simply replace the mask or delta
module but continue to follow the same retrieval recipe. 

\paragraph{Why retrieval-only persists.}
Even though these operators inherit limitations of the underlying
embedding, their \emph{tooling footprint} is minimal: no large
re-ranking LLM, no diffusion fallback, just one extra matrix multiply
before the ANN probe.  That cost-latency trade-off keeps them the
de-facto standard in production search today.

In the following, we categorize several of these key challenges.

\begin{enumerate}[label=\textbf{\arabic*.}, leftmargin=2em]

\item \textbf{Lack of labels for all visual attributes.}%
\newline
Many visual attributes lack explicit labels, making them hard to capture through standard learning approaches. While category labels support supervised tasks, abstract properties like style, fit, or texture (e.g., roughness, fluffiness) are rarely annotated and often resist discrete labeling, limiting the capacity of both discriminative and representation learning techniques to model them effectively.



    \item \textbf{Dataset Bias (Selection Bias)} \\
    Fashion data is often imbalanced: certain categories (like ``jeans'' or ``T-shirts'') dominate. Rare categories or attributes (like ``special pockets'' or ``rare designer prints'') get underrepresented, causing the model to ignore them.

    Oversampling or weighting can help, but does not always fix the deeper representation issues~\cite{Wang_2019_ICCV}.

    \item \textbf{Fast Emergence of New Fashion Concepts} \\
    New trends appear each season (``puff sleeves,'' ``cut-out dresses,'' ``shackets''), forcing re-labeling or retraining if we want the embedding to explicitly capture them.

    \item \textbf{Lack of Understanding of Negation or Compositionality} \\
    Queries like ``same shape, but no stripes'' or ``change color to navy while keeping everything else'' require more nuanced transformations than a static embedding typically provides.

    Some universal models conflate similar shapes with or without stripes, failing to separate them in a way that precisely handles the user’s text modification.
\end{enumerate}

\begin{figure}
    \centering    
    \includegraphics[width=0.70\linewidth]{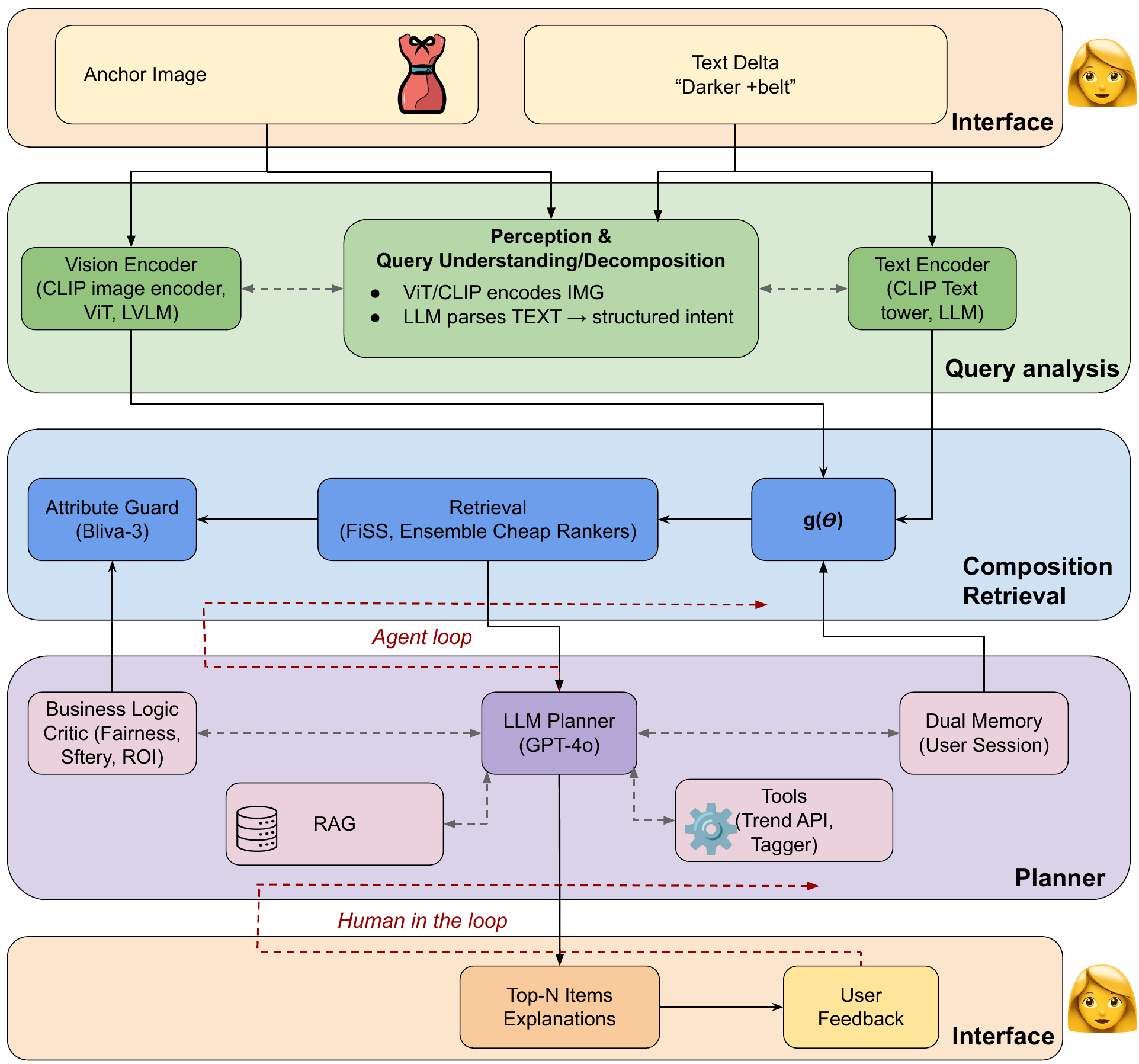}
 \caption{The four layers illustrate latency-critical (solid arrows) and reasoning paths (dashed arrows). User inputs (image + text) are embedded, composed by $g_{\theta}$, and queried via ANN. Candidates undergo attribute filtering, GPT-4o planner re-ranking with memory and tools, critic evaluation, and final rationale generation.}
 \label{fig:ammr}
\end{figure}

 
\section{Agentic Mixed-Modality Refinement (AMMR)}
\label{sec:ammr}
The AMMR pipeline integrates multimodal understanding, adaptive retrieval mechanisms, and agentic capabilities to address fundamental limitations of retrieval-only systems. It is structured into four layers—\textbf{Interface}, \textbf{Query Analysis}, \textbf{Composition Retrieval}, and \textbf{Planner}—with clearly defined interactions among components (Figure \ref{fig:ammr}).

\subsection{Interface and Query Analysis Layer}
At the top of the architecture, the interface allows users to express queries in a naturally multimodal fashion. Queries typically consist of an \textbf{anchor image} and a \textbf{text delta} (e.g., \textit{"darker + belt"}), offering users intuitive flexibility to express nuanced preferences.

The query analysis layer handles perceptual understanding and semantic decomposition through two encoders: a \textbf{Vision Encoder} (CLIP, ViT, LVLM) converting images into embeddings, and a \textbf{Text Encoder} (CLIP text tower, LLM) parsing textual input into structured constraints. An LLM planner translates colloquial expressions (e.g., \textit{"give me Bridgerton vibes"}) into structured vocabulary (\textsc{style\:cottagecore}), enhancing interpretability.

\subsection{Composition Retrieval and Agentic Planning Layers}
This section outlines the compositional retrieval and agent planning layers, which constitute the backbone of our approach to mixed-modality refinement.

\subsubsection{Composition Retrieval Layer}
This layer comprises:
\begin{enumerate}
\item \textbf{Composer ($ g_\theta$):} This module is aimed at dynamically fusing image embeddings and structured text constraints into a unified query vector, adapting similarity dimensions based on user session memory.
\item \textbf{Retrieval (FiSS, Ensemble Cheap Rankers):} Employs ensemble rankers that boost recall for underrepresented attributes by adaptive re-weighting.
\item \textbf{Attribute Guard (Bliva-3):} Verifies fine-grained attribute compliance post-retrieval, minimizing false positives.
\end{enumerate}

\textbf{What is the role of $g_{\theta}$?}

Operationally, $g_{\theta}$  acts as a query composer that maps multimodal inputs (image vector $v$, textual constraints $t$) into a composed query vector $q$. Architecturally, there are several viable options:
\begin{itemize}
\item \textbf{Gated-FiLM MLP (few-µs):} $q = W\_0 v + \sigma(W\_1 t) \odot v$, providing efficient fusion and gating mechanisms.
\item \textbf{Slice-wise $\Delta$-shift:} partitioning $v = [c \parallel m \parallel s \parallel \dots]$ and adding text-conditioned deltas only to the relevant subspaces.
\item \textbf{Memory-conditioned composer:} incorporating session memory keys into composer weights, allowing $g_\theta$ to adapt dynamically—for example, reducing the activation of floral features if consistently rejected by the user.
\end{itemize}

Note that the composed multimodal query vector $q$, produced by composer $g_{\theta}$, is efficiently queried against a nearest-neighbor index (e.g., FAISS) to retrieve an initial candidate set of 200–500 items. These candidates undergo immediate lightweight attribute verification (e.g., via BLIP-2) to enforce constraints such as color, price, gender, and brand-specific rules, significantly reducing computational overhead. Compared to static embeddings, this method dynamically composes queries, better capturing nuanced user intent.

\subsubsection{Agentic Planning Layer}

A GPT-4o-based agent orchestrates a structured reasoning cycle:
\begin{itemize}
\item \textbf{Thought:} Parses user constraints, identifying relevant tools or APIs (e.g., trend databases, brand rules).
\item \textbf{Action:} Dynamically invokes appropriate external tools or attribute-specialist rankers, refining candidate lists.
\item \textbf{Critic:} Evaluates recommendations for safety, fairness, and ROI, eliminating unsuitable options.
\item \textbf{Speak:} Provides ranked recommendations accompanied by concise, transparent rationales, enhancing user trust.
\end{itemize}

Overall, this structured reasoning process enables the system to flexibly incorporate external knowledge, adapt to evolving user intent in real time, and ensure recommendations remain both contextually relevant and aligned with multi-stakeholder objectives.

\subsection{Addressing Retrieval-Only Limitations}
Table \ref{tab:retrieval_limitations_AMMR} summarises the one-to-one mapping between classical bottlenecks and their AMMR counterparts. Two patterns are worth stressing. First, every limitation that stems from a static representation (unseen attributes, frozen similarity dimensions) is countered by a dynamic module: either the memory-conditioned composer $g_{\theta}$ or the ensemble rankers whose weights are re-learnt from user feedback. Second, language vagueness (colloquialisms, negation) is never handled inside the vector space itself but delegated to the LLM planner, which rewrites the query into explicit, machine-verifiable slots before search. This separation keeps the ANN index fast while still achieving semantic coverage that static retrieval alone could not offer.
\begin{table}[ht]
\centering
\caption{Retrieval-only limitations vs. proposed AMMR solutions.}
\label{tab:retrieval_limitations_AMMR}
\renewcommand{\arraystretch}{1.5}
\begin{tabular}{@{}p{0.43\linewidth}@{\hspace{3pt}}c@{\hspace{3pt}}p{0.48\linewidth}@{}}
\toprule
\textbf{Retrieval-only limitation} & & \textbf{AMMR solution} \\
\midrule
Fine attributes unseen (\(\Delta\)) 
& $\rightarrow$ 
& Pool of {attribute-specialist rankers} + {attribute verifier} boosts recall on \textit{tail attributes}
\\
\addlinespace[4pt]

Colloquial or elliptical queries (\dquote{\textit{give me Bridgerton vibes}})
& $\rightarrow$ 
& LLM Planner maps colloquialisms to controlled vocab (\textsc{style:cottagecore}) before search \\
\addlinespace[4pt]

Rapid trend drift 
& $\rightarrow$ 
& Planner accesses \textbf{external trend API}; Memory injects recent style tokens into composer \( g_\theta \). \\
\addlinespace[4pt]

Negation \& compositionality (“no stripes”) 
& $\rightarrow$ 
& LLM interprets negation, rewriting text deltas; composer subtracts specific attribute activation vectors. \\
\addlinespace[4pt]

Single-notion similarity dominates (colour $\gg$ fit) 
& $\rightarrow$ 
& Ensemble rankers re-weight dimensions per session. \\
\bottomrule
\end{tabular}
\end{table}

\subsection{Open Challenges}
The following seven questions capture the most actionable gaps:
\begin{itemize}
    \item \textbf{Efficient Adaptive Composer.} How can we design an adaptive composer $g_{\theta}$ that warps similarity spaces online while respecting a strict per-query GPU budget?

    \item \textbf{Privacy-Preserving Session Memory.} How can personalised session memory be scaled—e.g., via on-device vector stores and federated distillation—without compromising user privacy?

    \item \textbf{Multi-Objective Agentic Critic.} What real-time algorithms can reconcile safety, user \& producer fairness, ROI, and platform revenue in multi-agent recommender settings~\cite{maragheh2025future}?

    \item \textbf{RAG-Enhanced Verification.} Which retrieval-augmented vision–language architectures (e.g., RAG-VisualRec) can achieve high tail-attribute recall with minimal human labels~\cite{tourani2025rag}?

    \item \textbf{Holistic Evaluation Protocol.} How can we fuse offline, online, user-centric, and environmental signals into a single Pareto frontier for fashion agents~\cite{deldjoo2025toward}?

    \item \textbf{Safe Data Augmentation.} Which generation pipelines—such as diffusion outfit synthesis—can expand long-tail coverage without degrading signal-to-noise or brand coherence~\cite{banerjee2025synthtrips}?

    \item \textbf{Agentic Query Negotiation.} How can an agent proactively detect under-specified requests, elicit concise clarifications in multi-turn dialogue, and still meet real-time latency and privacy constraints?
\end{itemize}

\section{Conclusion}
Traditional retrieval-only pipelines often fall short in meeting the nuanced demands of contemporary fashion recommendation, particularly when faced with intricate visual attributes and multifaceted user queries. In response, we introduce the Agentic Multimodal Modular Recommender (AMMR), a holistic framework that unites multimodal encoders, adaptive query composition, dynamic retrieval mechanisms, and agentic planning modules powered by large language models (LLMs). This generative, multimodal, and agentic approach addresses several previously unfulfilled requirements in fashion recommendation: capturing fine-grained and long-tail visual attributes, resolving ambiguous and compositional user queries, and adapting in real time to evolving trends and contexts. Moreover, AMMR enables inherently explainable recommendation processes, thereby fostering greater user trust, transparency, and satisfaction.

Despite its demonstrated potential, AMMR also surfaces several open challenges that warrant further research attention. These include the development of computationally efficient and resource-constrained adaptive composer modules, the design of privacy-preserving personalization strategies (e.g., through on-device memory or federated learning), and the reconciliation of conflicting multi-objective constraints such as fairness, safety, return on investment, and platform revenue within an agentic multi-stakeholder environment. In addition, safeguarding against hallucinations in LLM-generated explanations, ensuring robust retrieval-augmented verification, deploying safe and effective data augmentation pipelines, and developing holistic evaluation protocols that reflect both system-level and user-centric metrics remain crucial.

Addressing these open challenges will not only push the frontiers of generative and agentic recommender systems, but also enable the realization of truly adaptive, interactive, and trustworthy fashion stylists for the next era of personalized recommendation.

\bibliographystyle{ACM-Reference-Format}
\bibliography{refs}

\end{document}